\documentclass[pdflatex,sn-mathphys-num]{sn-jnl}

\usepackage{graphicx}%
\usepackage{multirow}%
\usepackage{amsmath,amssymb,amsfonts}%
\usepackage{amsthm}%
\usepackage{mathrsfs}%
\usepackage[title]{appendix}%
\usepackage{xcolor}%
\usepackage{textcomp}%
\usepackage{manyfoot}%
\usepackage{booktabs}%
\usepackage{algorithm}%
\usepackage{algorithmicx}%
\usepackage{algpseudocode}%
\usepackage{listings}%

\theoremstyle{thmstyleone}%
\theoremstyle{thmstyletwo}%

\theoremstyle{thmstylethree}%

\begin{document}

\title[Article Title]{Peltier cooling in Corbino-geometry quantum Hall systems}


\author*[1]{\fnm{Akira} \sur{Endo}}\email{akrendo@issp.u-tokyo.ac.jp}

\author[1]{\fnm{Yoshiaki} \sur{Hashimoto}}


\affil*[1]{\orgdiv{Institute for Solid State Physics}, \orgname{The University of Tokyo}, \orgaddress{\street{Kashiwanoha}, \city{Kashiwa}, \postcode{277-8581}, \state{Chiba}, \country{Japan}}}




\abstract{Quantum Hall systems having Corbino geometry are expected to have a large Peltier coefficient $\Pi_{rr}$ in the quantum Hall plateau region. We present an analytic formula for $\Pi_{rr}$ calculated employing the spectral conductivity obtained based on the self-consistent Born approximation.  The coefficient $\Pi_{rr}$ is shown to have a large negative (positive) value just above (below) an integer Landau-level filling, with the absolute value $|\Pi_{rr}|$ increasing with decreasing temperature or decreasing disorder, and approaching the saw-tooth shape $- (E_{N_\mathrm{F} \sigma_\mathrm{F}}-\zeta)/e$ in the limit of vanishing disorder, where $E_{N_\mathrm{F} \sigma_\mathrm{F}}$ is the highest occupied Landau level and $\zeta$ is the chemical potential. As an initial attempt to experimentally observe the effect of the large $|\Pi_{rr}|$, we measure the electron temperature $T_\mathrm{out}$ near the outer perimeter of a Corbino disk, applying a radial dc current $I_\mathrm{dc}$. The temperature $T_\mathrm{out}$ is observed to increase or decrease depending on the direction of $I_\mathrm{dc}$ and the sign of $\Pi_{rr}$ as expected from the Peltier effect. Notably, $T_\mathrm{out}$ becomes lower than the bath temperature for outward (inward) $I_\mathrm{dc}$ in the region where $\Pi_{rr} < 0$ ($\Pi_{rr} > 0$).}




\maketitle

\section{Introduction}\label{sec1}
The lack of dissipationless edge channels connecting inner and outer electrodes of a Corbino disk vastly alters the transport properties of the quantum Hall systems (two-dimensional electron system, 2DES, subjected to a quantizing magnetic field) compared to their counterparts in a Hall-bar device. An archetypal example of much relevance to the present study is the thermoelectric coefficients\footnote{We neglect the contribution of phonons throughout the paper, which is justified at very low temperatures where phonons die out.}. While the (diagonal) Seebeck coefficient $S_{xx}$ vanishes in the quantum Hall plateau regions in Hall-bar devices \cite{Jonson84,Ying94}, the (radial) Seebeck coefficient $S_{rr}$ of a Corbino sample is known to have a large negative (positive) value in the quantum Hall plateau region just above (below) an integer Landau-level filling, where thermal flux is carried by electrons (holes) \cite{Barlas12,Kobayakawa13,Real20}.
The difference in the Seebeck coefficient between the two configurations is reminiscent of that in the resistivity, with the diagonal resistivity $\rho_{xx}$ in the Hall-bar geometry and the radial resistivity $\rho_{rr}$ in the Corbino geometry becoming vanishingly small and divergently large, respectively, in the quantum Hall plateau regions. In fact, the differences seen in the two geometries can be traced back to the common origin: the lack of the contribution from the off-diagonal component of the electric conductivity in the Corbino geometry. Derivation of $S_{xx}$ and $S_{rr}$ from transport equations was detailed in Refs.\ \cite{Barlas12,Kobayakawa13}.
Further insight on the Corbino Seebeck coefficient $S_{rr}$ was provided in Ref.\ \cite{Barlas12}, which showed that $S_{rr}$ can be interpreted as \textit{the entropy per carrier per carrier charge}. The increase of $|S_{rr}|$ when approaching the integer fillings or reducing the temperature [see Fig.\ \ref{Tdeps}(b)] can be interpreted as resulting from the decrease in the number of the carriers.

Seen from a wider perspective, the large thermoelectric response of the Corbino quantum Hall systems has much in common with remarkable thermoelectricity in other systems composed of gapped extended states and localized states within the gap \cite{Guttman95,Yamamoto17,Dey22}. Note that the dissipationless edge states shortcircuits the localized state in the quantum Hall plateau regions in the Hall-bar devices, nullifying the role played by localized states. By contrast, the lack of the edge states connecting the relevant electrodes allows a Corbino device in the quantum Hall plateau regions to behave like other systems with localized states. In the GaAs/AlGaAs 2DES considered in the present study, the gap is generated by the Landau quantization and the spin splitting in even-integer and odd-integer quantum Hall states, respectively. We consider spins in this study only to introduce the odd-integer energy gaps, which are generally smaller than the even-integer energy gaps. We do not examine spin-specific transport phenomena as in the spin caloritronics \cite{BauerRev12,UchidaRev21,YangRev23,Uchida08,Benenti17,Trocha22,Trocha25}.

According to the Kelvin-Onsager relation $\Pi_{rr} = T S_{rr}$ \cite{ZimanEP60}, the (radial) Peltier coefficient $\Pi_{rr}$ is also expected to take a large value in the quantum Hall plateau region in Corbino devices. In the present study, we focus on the large $\Pi_{rr}$ in the Corbino quantum Hall systems. To gain an overview of the magnitude of $\Pi_{rr}$ and to see how it varies with temperature and disorder, we calculate $\Pi_{rr}$ for typical experimental parameters employing an approximate analytic formula deduced from the spectral conductivity $\sigma_{0,rr}$ (electric conductivity at zero temperature) obtained based on self-consistent Born approximation (SCBA) \cite{AndoLB74}. We find that $|\Pi_{rr}|$ increases with decreasing temperature and decreasing disorder. We also present a simple formula representing the upper limit of the magnitude of $\Pi_{rr}$. In search of the experimental evidence for the large $|\Pi_{rr}|$, we investigate the response of the electron temperature $T_\mathrm{our}$ near the outer perimeter of a Corbino disk to the radial dc current $I_\mathrm{dc}$, employing the capacitance between the 2DES and an annular top gate placed near the outer perimeter as the measure of $T_\mathrm{out}$. The temperature $T_\mathrm{out}$ thus observed is found to either increase or decrease depending on the direction of $I_\mathrm{dc}$ and whether the Landau-level filling fraction is above or below an integer value. The observed variation in $T_\mathrm{out}$ is ascribable to the thermal flux carried by $I_\mathrm{dc}$ due to the Peltier effect.

\section{Calculation of the Peltier coefficient}\label{secCalcPi}
We first deduce an analytic formula for the Peltier coefficient $\Pi_{rr}$. Although based on rather crude approximations as will be described below, it is still useful in grasping how $\Pi_{rr}$ varies with the temperature or disorder. We start from the electric conductivity at $T = 0$ K, $\sigma_{0,rr}$, obtained by the self-consistent Born approximation (SCBA) \cite{AndoLB74}. The SCBA is selected mainly because it yields analytic expressions for the Seebeck coefficient $S_{rr}$ and $\Pi_{rr}$. Although the semi-elliptical disorder-broadened Landau levels (LLs) resulting from SCBA are at variance with experimentally observed LLs, which are better represented by Gaussian or Lorentzian\cite{Gornik85,Berendschot87,Ashoori92,Dial07,Eisenstein85,Potts96,Zhu03,Endo08SdHH}, the electric conductivity $\sigma_{rr}$ and $S_{rr}$ deduced by SCBA reproduce the experimental observations fairly well \cite{Kobayakawa13}.

Modifying the expression neglecting the spin, given by Eqs.\ (3.6) and (3.10)\footnote{Equation (3.10) in Ref.\ \cite{AndoLB74} assumes short-range scatterers, which is not necessarily appropriate for the GaAs/AlGaAs 2DES \cite{Coleridge91}. We employed this approximation just for simplicity.} in Ref.\ \cite{AndoLB74}, to incorporate the spin splitting,\footnote{This is done, without strict theoretical underpinnings, simply by replacing the spinless LLs $E_N$ with the LLs with spin $E_{N\sigma}$. This modification is necessary to introduce odd-integer quantum Hall states and again, describes experimental observations fairly well.} we have
\begin{equation}
\sigma_{0,rr}\left(E\right)
= \frac{e^2}{h}\frac{2}{\pi}\ \sum_{N \sigma}\left(N+\frac{1}{2}\right)\mathrm{Max} \left\{1-\left(\frac{E-E_{N\sigma}}{\Gamma}\right)^2,0\right\} ,
\label{Eqsgm0}
\end{equation}
where
\begin{equation}
E_{N \sigma}=\left(N+\frac{1}{2}\right)\hbar\omega_\mathrm{c} +\frac{1}{2} \sigma g^* \mu_\mathrm{B} B
\end{equation}
is the spin-resolved $N$-th Landau level with $\omega_\mathrm{c} = e B / m^*$ the cyclotron angular frequency, $m^*$ the electron effective mass, and $\sigma = \pm 1$ the spin index. 
The effective g-factor $g^*$ varies with the Landau-level filling fraction $\nu = n_\mathrm{e} h / eB$ due to the exchange enhancement at odd-integer fillings \cite{AndoOG74}, where $n_\mathrm{e}$ is the electron density. The model for $g^*$ used in the present calculation is detailed in Appendix \ref{secAPXg}. 
Disorder-broadened width of the Landau level
\begin{equation} 
\Gamma = \frac{e\hbar}{m^*} \sqrt{\frac{2 B}{\pi \mu_\mathrm{q}}}  \label{Gamma}
\end{equation}
is related to the quantum mobility $\mu_\mathrm{q}$, which can be experimentally deduced from the damping of the Shubnikov-de Haas oscillations \cite{Coleridge91}. 
The Seebeck coefficient $S_{rr}$ is related to the electric conductivity
\begin{equation}
\sigma_{rr}  = \int_{-\infty}^{\infty}\left(-\frac{\partial f}{\partial E}\right)\sigma_{0,rr}\left(E\right) dE \label{Eqsgmrr}
\end{equation}
and the thermoelectric conductivity\footnote{Here, we followed the notation adopted in \cite{Fletcher99,Endo23}}
\begin{equation}
\varepsilon_{rr} = -\frac{1}{eT}\int_{-\infty}^{\infty}\left(E-\zeta\right)\left(-\frac{\partial f}{\partial E}\right)\sigma_{0,rr}\left(E\right) dE \label{Eqepsrr}
\end{equation}
as
\begin{equation}
S_{rr} = \frac{\varepsilon_{rr}}{\sigma_{rr}}, \label{EqSrr}
\end{equation}
where $f\left(E\right) = \{1+\exp{[(E-\zeta)/(k_\mathrm{B}T)]}\} ^{-1}$ is the Fermi-Dirac distribution function with $\zeta$ the chemical potential. In the present study, we use a fixed value $\zeta = \pi \hbar^2 n_\mathrm{e} / m^* \equiv \zeta_0$ determined from the electron density $n_\mathrm{e}$ at $B =0$ T and $T \rightarrow 0$ K for simplicity, neglecting the possible dependence of $\zeta$ on $B$. The appropriateness of this approximation will be discussed at end of this section.
We further calculate $\Pi_{rr}$ from $S_{rr}$ with the Kelvin-Onsager relation
\begin{equation}
\Pi_{rr} = TS_{rr} \label{KelvOns}
\end{equation}
mentioned above. 
The calculation of $S_{rr}$ from SCBA $\sigma_{0,rr}$ has been performed numerically before \cite{Barlas12,Kobayakawa13}. In the present study, we take a step further and present analytic expressions for $\sigma_{rr}$ and $\varepsilon_{rr} $ from which to calculate $S_{rr}$ and $\Pi_{rr}$. 
By performing the integral in Eqs. (\ref{Eqsgmrr}) and (\ref{Eqepsrr}), we obtain 
\begin{equation}
\sigma_{rr} = \frac{e^2}{h}\frac{2}{\pi}
\sum_{N \sigma}\left(N+\frac{1}{2}\right)\left[\alpha_{N \sigma} D_0(\xi_{N \sigma}, \eta)+\beta_{N \sigma}  D_1(\xi_{N \sigma}, \eta)-\gamma D_2(\xi_{N \sigma}, \eta)\right]
\label{EqsgmrrSCBA}
\end{equation}
and
\begin{equation}
\varepsilon_{rr} = -\frac{k_B}{e}\frac{e^2}{h}\frac{2}{\pi}
\sum_{N \sigma}\left(N+\frac{1}{2}\right)\left[\alpha_{N \sigma} D_1(\xi_{N \sigma}, \eta)+\beta_{N \sigma} D_2(\xi_{N \sigma}, \eta)-\gamma D_3(\xi_{N \sigma}, \eta)\right],
\label{EqepsrrSCBA}
\end{equation}
respectively, where
\begin{subequations}
\begin{align}
\alpha_{N \sigma} & \equiv 1- \left( \frac{E_{N \sigma}-\zeta}{\Gamma} \right)^2, \label{Eqalpha} \\
\beta_{N \sigma} & \equiv 2 \frac{k_\mathrm{B} T}{\Gamma}  \frac{E_{N \sigma}-\zeta}{\Gamma}, \label{Eqbeta} \\
\gamma & \equiv \left( \frac{k_\mathrm{B} T}{\Gamma} \right) ^2, \label{Eqgamma}
\end{align}
\end{subequations}
\begin{subequations}
\begin{align}
\xi_{N \sigma} & \equiv \frac{E_{N \sigma}-\zeta}{k_\mathrm{B} T}, \label{xinsgm} \\
\eta & \equiv \frac{\Gamma}{k_BT}, \label{eta}
\end{align}
\end{subequations}
and
\begin{subequations} \label{EqDandA}
\begin{align} 
& \hspace{10mm} D_n(\xi, \eta) \equiv A_n(\xi+\eta)-A_n(\xi-\eta), \label{EqD} \\
& A_n\left(x\right) \equiv \int  x^n \left( -\frac{\partial \tilde{f}}{\partial x} \right) d x = -x^n  \tilde{f} \left(x\right)+nF_n\left(x\right),  \label{EqA} \\
& \hspace{55mm} (n = 0, 1, 2, 3) \nonumber
\end{align}
\end{subequations}
with
\begin{subequations}
\begin{align}
\tilde{f} (x) & \equiv \frac{1}{1+\exp x}, \label{tildef} \\
F_n\left(x\right) & \equiv\int d x x^{n-1} \tilde{f} \left(x\right). \label{EqF} 
\end{align}
\end{subequations}
The function $F_n(x)$ can be written analytically using polylogarithm functions as shown in Appendix \ref{secAPXF} for $n = 1 -3$.

\begin{figure}[h]
\centering
\includegraphics[width=1.0\textwidth]{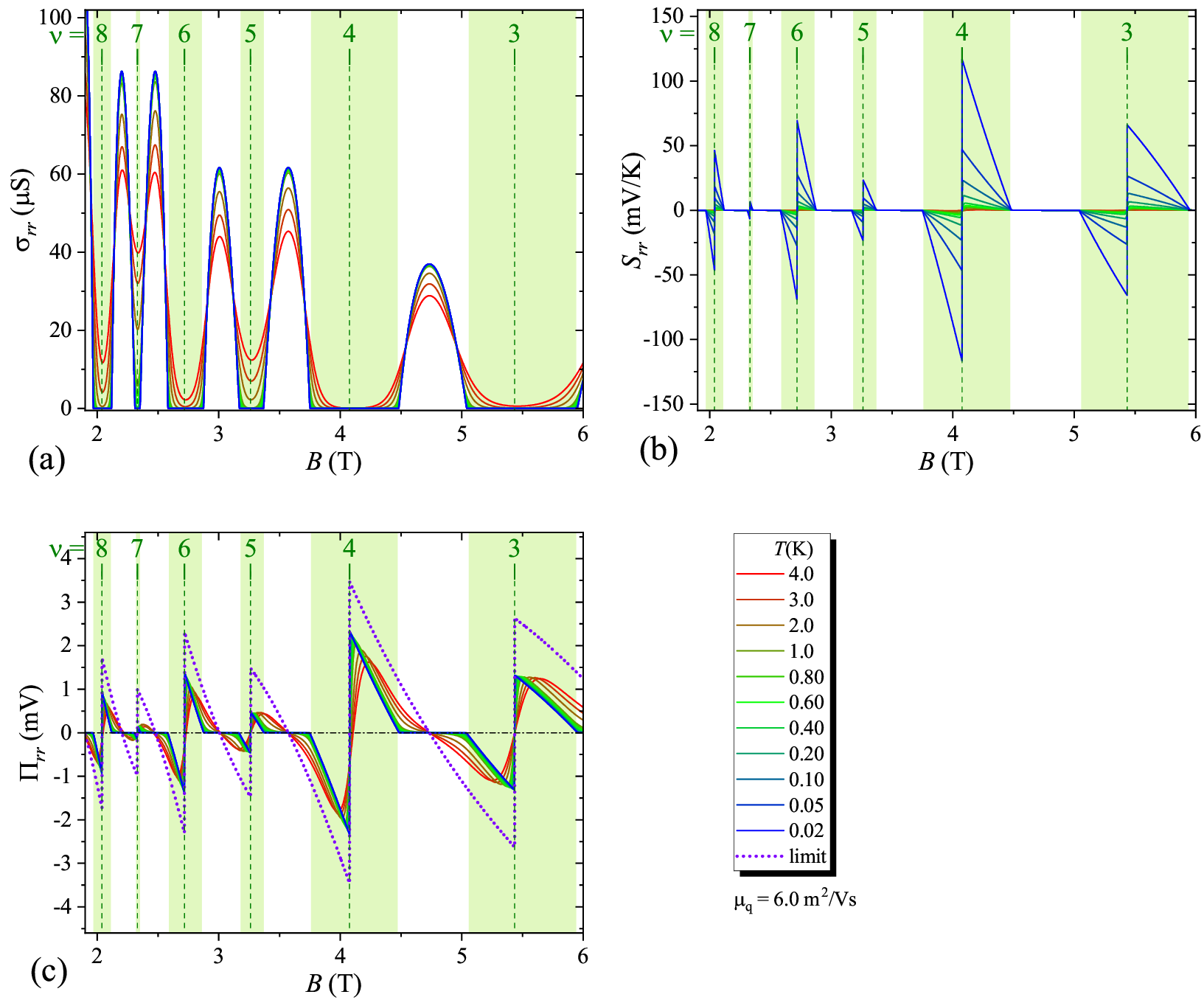}
\caption{Magnetic-field dependence of the conductivity $\sigma_{rr}$ (a), the Seebeck coefficient $S_{rr}$ (b), and the Peltier coefficient $\Pi_{rr}$ (c) for various temperatures and the quantum mobility $\mu_\mathrm{q} = 6.0$ m$^2$/Vs. Green shaded areas highlight the quantum Hall plateau regions (for $T= 0.20$ K) and the vertical dashed lines with numbers mark the location of exact integer Landau-level fillings. The dotted line in (c) shows the upper limit of $|\Pi_{rr}|$ given by Eq.\ (\ref{EqPirrCLT}).}\label{Tdeps}
\end{figure}

In Fig.\ \ref{Tdeps}, we plot $\sigma_{rr}$, $S_{rr}$, and $\Pi_{rr}$ calculated for various temperatures ranging from 4.0 K to 0.02 K in the magnetic-field range $1$ T $\leq B \leq 6$ T\@. The sample parameters were taken from the 2DES used in our experiment: $n_\mathrm{e} = 3.94\times 10^{15}$ m$^{-2}$ and $\mu_\mathrm{q} = 6.0$ m$^2$/Vs, determined from the frequency and the damping amplitude of the SdH oscillations, respectively, and $g^*$ described in Appendix \ref{secAPXg}. 
We basically used Eqs.\ (\ref{EqsgmrrSCBA}) and (\ref{EqepsrrSCBA}) for the calculation. In the close vicinity of integer fillings, however, we resorted to the approximation described in Appendix \ref{secAPXP} and replaced $D_n(\xi,\eta)$ in the equations with the approximate formulas Eq.\ (\ref{ADs}).  
The green shaded areas in the figures highlight the quantum Hall plateau regions, and the locations of the integer fillings are indicated by vertical dashed lines. We can see that both $S_{rr}$, and $\Pi_{rr}$ have a large value in the quantum Hall plateau regions, enhancing its magnitude toward exact integer fillings. The absolute value $|S_{rr}|$ rapidly grows with decreasing temperature. The growth is roughly proportional to $1/T$.  Accordingly, the variation of $|\Pi_{rr}|$ with the temperature is relatively small, but the growth toward the integer fillings still becomes more pronounced with decreasing temperature.  

\begin{figure}[h]
\centering
\includegraphics[width=1.0\textwidth]{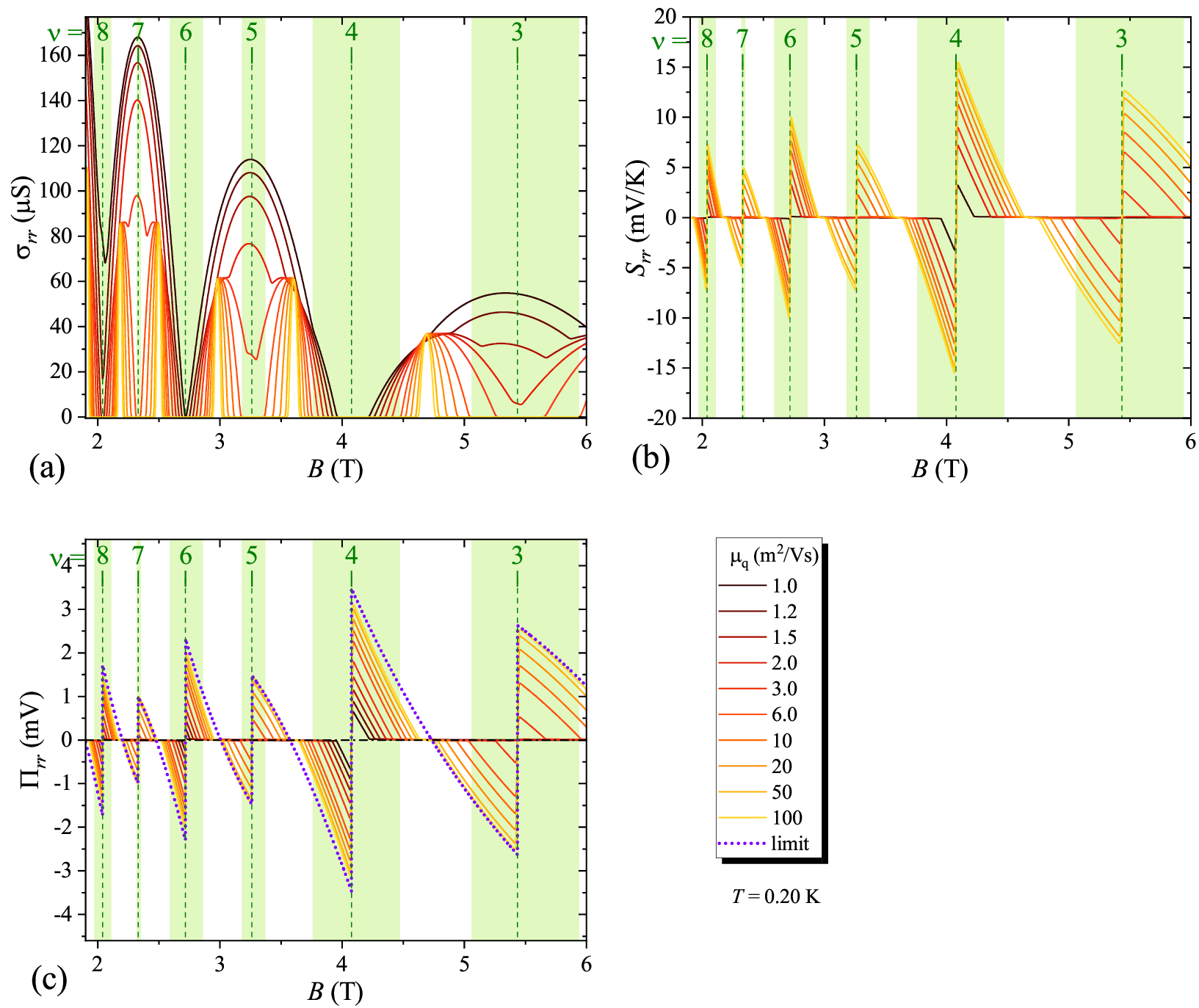}
\caption{Magnetic-field dependence of the conductivity $\sigma_{rr}$ (a), the Seebeck coefficient $S_{rr}$ (b), and the Peltier coefficient $\Pi_{rr}$ (c) for various quantum mobilities at $T = 0.20$ K\@. Green shaded areas highlight the quantum Hall plateau regions (for $\mu_\mathrm{q} = 6.0$ m$^2$/Vs) and the vertical dashed lines with numbers mark the location of exact integer Landau-level fillings. The dotted line in (c) shows the upper limit of $|\Pi_{rr}|$ given by Eq.\ (\ref{EqPirrCLT}).}\label{Qdeps}
\end{figure}

To illustrate the effect of disorder, we repeated the calculation for $T = 0.20$ K with $\mu_\mathrm{q}$ replaced with various values ranging from 1.0 m$^2$/Vs to 100 m$^2$/Vs. The result, plotted in Fig.\ \ref{Qdeps}, reveals that both $|S_{rr}|$ and $|\Pi_{rr}|$  increase with decreasing disorder (increasing $\mu_\mathrm{q}$), and $\Pi_{rr}$ approaches that of the low-disorder limit [Eq.\ (\ref{EqPirrCLT})] deduced in Appendix \ref{secAPXD}.

In the above calculations, we approximated the chemical potential $\zeta$ by a constant value $\zeta_0 = \pi \hbar^2 n_\mathrm{e} / m^*$. Unless the carrier density is allowed to vary with $B$, however, $\zeta$ should oscillate with $B$ around $\zeta_0$, approaching the topmost occupied Landau level $E_{N_\mathrm{F}\sigma_\mathrm{F}}$ 
(see, e.g., \cite{Nizhankovskii86,Dabiran88,Endo08SdHH}). Consequently, $|S_{rr}|$ and $|\Pi_{rr}|$ will become smaller due to the reduction in $|\xi_{N\sigma}|$. However, the approach becomes less pronounced and the amplitude of the oscillations is kept smaller for a more disordered 2DES with larger $\Gamma$. For the parameters of the sample used in the present study, the oscillation amplitude is at most about 10\% of $\zeta_0$. Furthermore, the $B$-dependent $\zeta$ crosses $\zeta_0$ at integer filings and remains close to $\zeta_0$, pinned by the localized states, in the vicinity of the integer fillings, where we place the main focus in this study. The constant $\zeta$ thus remains a fairly good approximation in the present study.

\section{Experimental evidence for the Peltier effect}\label{secExp}
\begin{figure}[h]
\centering
\includegraphics[width=1.0\textwidth]{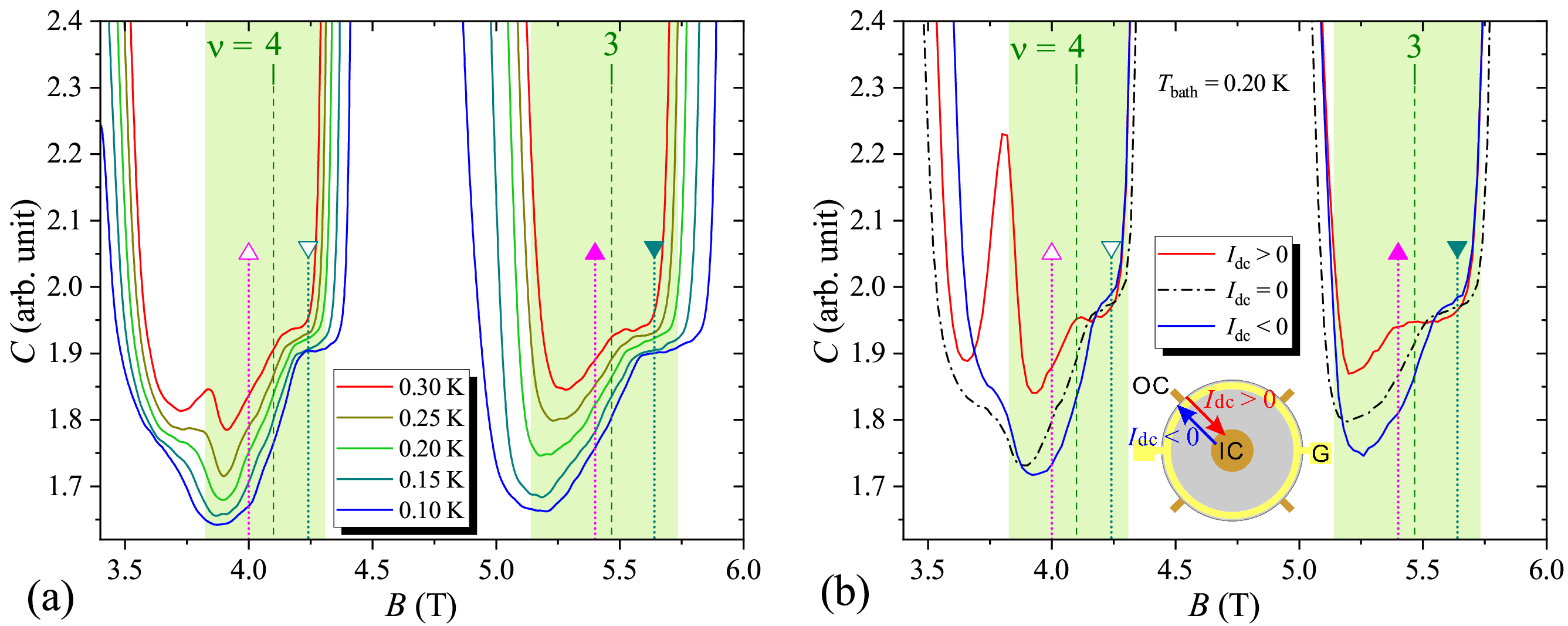}
\caption{Experimentally measured magnetic-field dependence of the capacitance $C$ between the annular top gate and the 2DES measured at various bath temperatures $T_\mathrm{bath}$ (a) and at $T_\mathrm{bath} = 0.20$ K with inward ($I_\mathrm{dc} > 0$) or outward ($I_\mathrm{dc} < 0$) radial dc current, or without the dc current ($I_\mathrm{dc} = 0$) (b). The green shaded areas indicate the quantum Hall plateau areas, with vertical dashed lines marking the locations of exact integer fillings. Upward and downward triangles with vertical dotted lines indicate the positions of the magnetic field selected for further examination shown in Fig.\ \ref{FigTCD} and Table \ref{table}.  Inset in (b) depicts the schematics of the measurement device. G: annular top gate. IC: inner electrode. OC: outer electrode.}\label{expFigs}
\end{figure}

In the previous section, we have seen that the Peltier coefficient of the Corbino-shaped quantum Hall systems can become extremely large in the quantum Hall plateau regions. As an initial attempt to experimentally observe the evidence for the large Peltier effect, we measured the response of the electron temperature $T_\mathrm{out}$ near the outer rim of a Corbino disc to the radial dc current $I_\mathrm{dc}$. The temperature  measurement has to be done without disturbing the 2DES temperature. To this end, we employed the capacitance between the top gate and the 2DES as a measure of the electron temperature. Since the top gate does not have direct electrical contact to the 2DES, we expect it has minimal effect on the electron temperature.

\begin{figure}[h]
\centering
\includegraphics[width=0.5\textwidth]{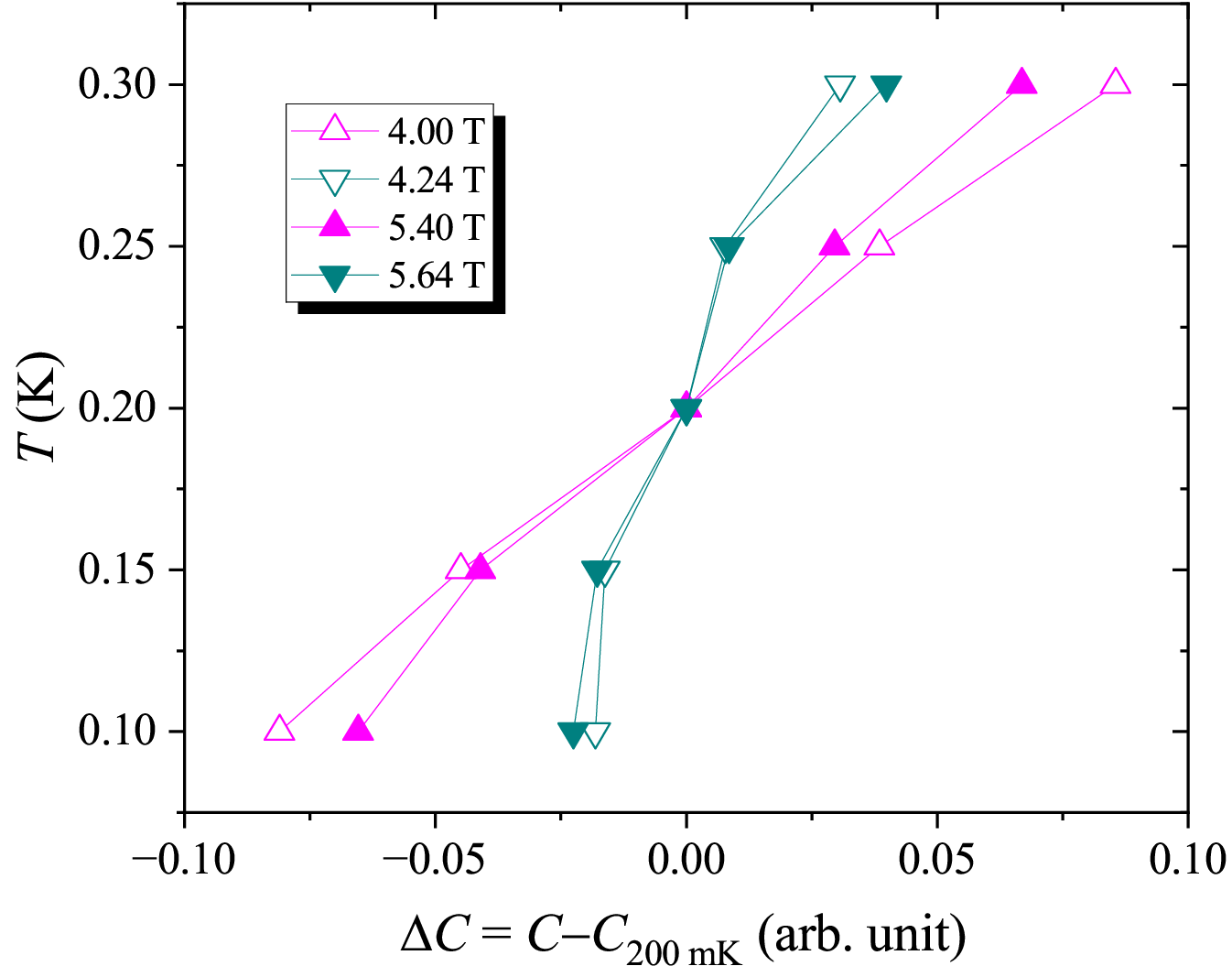}
\caption{Temperature $T$ versus the difference $\Delta C$ of the capacitance $C$ from that at $T = 0.20$ K picked out from Fig.\ \ref{expFigs}(a) at four magnetic fields in the quantum Hall plateau regions. The symbols for the plot here are the same as those employed to indicate the magnetic-field locations in Fig.\ \ref{expFigs}. }\label{FigTCD}
\end{figure}

The Corbino device used in the present study, with the radius 1 mm, was fabricated from a GaAs/AlGaAs 2DES wafer with  $n_\mathrm{e} = 3.94\times 10^{15}$ m$^{-2}$ and $\mu_\mathrm{q} = 6.0$ m$^2$/Vs as mentioned earlier, and the mobility $\mu = 78$ m$^2$/Vs. The device was immersed in the mixing chamber of the dilution refrigerator for the low temperature measurement. 
 As shown in the inset of Fig.\ \ref{expFigs}(b), an annular top gate was placed near the outer perimeter of the Corbino disk to measure the capacitance $C$ between the gate and the 2DES beneath it. It is well known that the capacitance $C$ between a top gate and a 2DES becomes much smaller in the quantum Hall plateau regions compared with $C$ in the regions between the quantum Hall states \cite{Goodall85,Oto95,Oto96}. Figure \ref{expFigs}(a) shows the capacitance $C$ measured around the plateau areas of the $\nu =3$ and 4 quantum Hall states for various temperatures $T_\mathrm{bath}$ of the bath in which the sample was immersed. The quantum Hall plateau areas with smaller $C$ are highlighted in the figure. It can readily be seen that $C$ increases with $T_\mathrm{bath}$ for the entire magnetic-field range shown in the figure. 
The mechanism responsible for the temperature dependence is not fully understood at present. We consider, however, that the temperature dependence mainly results from the temperature dependence of the electric conductivity of the 2DES beneath the annular top gate. The resistive plate model described in Refs.\ \cite{Oto95,Oto96} allows us to see that the temperature dependence of $C$ is consistent with that of the electric conductivity. 
In this study, we exploit this temperature dependence as a noninvasive thermometer and interpret the increase/decrease of $C$ as reflecting the increase/decrease of the electron temperature $T_\mathrm{out}$ around the outer rim of the Corbino disk.
Figure \ref{FigTCD} exemplifies the relation employed for the capacitance-to-temperature conversion. The figure plots the temperature versus the capacitance read out from Fig.\ \ref{expFigs}(a) at magnetic fields slightly above (upward triangles) and below (downward triangles) integer fillings $\nu = 4$ (open symbols) and $\nu = 3$ (solid symbols). The locations are marked in Fig.\ \ref{expFigs} using the same symbols.

In an attempt to observe the Peltier effect, we measured $C$ while applying a radial dc current $I_\mathrm{dc}$. In Fig.\ \ref{expFigs}(b), we plot $C$ measured at $T_\mathrm{bath} = 0.20$ K with outward ($I_\mathrm{dc} \simeq -1.5$ nA, blue line) and inward ($I_\mathrm{dc} \simeq 1.5$ nA, red line) dc current, and without the dc current ($I_\mathrm{dc} = 0$ nA, black dot-dashed line). In the regions with $\nu$ slightly larger than an integer within the plateau areas, where $\Pi_{rr} < 0$, $C$ decreases (increases) for $I_\mathrm{dc} < 0$ ($I_\mathrm{dc} > 0$) compared to $C$ for $I_\mathrm{dc} = 0$, suggesting that $T_\mathrm{out}$ has decreased (increased) from $T_\mathrm{bath}$. The change in the temperature is in line with the Peltier effect, which lets the thermal flux be carried inward (outward) by the outward (inward) $I_\mathrm{dc}$. The trend in the temperature change is reversed, although less apparent, in the regions with $\nu$ slightly smaller than an integer. Noting that $\Pi_{rr}$ changes sign ($\Pi_{rr} > 0$) in these regions, the change in the temperature is again consistent with the Peltier effect. The smaller change in $C$ for the latter regions is partly attributable to the smaller $T$-to-$C$ conversion ratio apparent in Figs.\ \ref{expFigs}(a) and \ref{FigTCD}. 
The reproducibility of the behavior seen in Fig.\ \ref{expFigs}(b) is demonstrated in Appendix \ref{secAPRpr}.
Table \ref{table} tabulates, as examples, the $I_\mathrm{dc}$-induced change $\Delta C$ in the capacitance read out from Fig.\ \ref{expFigs}(b) and the temperature estimated by translating $\Delta C$ to $T$ employing Fig.\ \ref{FigTCD}, assuming that the observed $\Delta C$ is fully attributable to the temperature change, for the selected four magnetic fields.
The possibility that $I_\mathrm{dc}$ affects $\Delta C$ through non-thermal purely electric mechanisms cannot be completely ruled out. The asymmetric behavior of $\Delta C$ above and below integer fillings can be caused by the superposition of such effects, leading to overestimation (underestimation) of the change of the temperature shown in Table 1 above (below) integer fillings. It is unlikely,  however, that $\Delta C$ due to pure electrical effects changes sign by crossing integer fillings, suggesting that such effects, if present, are outweighed by the Peltier effect. 
Remarkably, $T_\mathrm{out}$ can become lower than $T_\mathrm{bath}$ for appropriate choice of $\nu$ and the direction of $I_\mathrm{dc}$. Outside the plateau regions, by contrast, $C$ increases when $I_\mathrm{dc}$ is applied, regardless of the direction. In these areas, $\Pi_{rr}$ is small, and the Joule heating, which does not depend on the direction of $I_\mathrm{dc}$, outweighs the Peltier effect, resulting in the temperature increase.

\begin{table}[h]
\centering
\caption{Temperatures estimated from the change in the capacitance at the magnetic fields indicated by triangles in Fig.\ \ref{expFigs}. $T_\mathrm{bath} = 0.20$ K\@. \label{table}}
\begin{tabular}{cc|cc|cc}  \hline \hline
 & & \multicolumn{2}{c|}{$I_\mathrm{dc} > 0$} &  \multicolumn{2}{c}{$I_\mathrm{dc} < 0$} \\
$B$ (T) & $\nu$ & $\Delta C$ (arb. unit) &  $T$ (K)  & $\Delta C$ (arb. unit)  & $T$ (K)  \\ \hline
4.00 & 4.10 & 0.081 & 0.29 & $-$0.065 & 0.12 \\ 
4.24 & 3.87 & $-$0.044 & 0.19  & 0.016 & 0.27 \\ 
5.40 & 3.14 & 0.073 & 0.31 & $-$0.057 & 0.12 \\ 
5.64 & 2.91 & $-$0.004 & 0.19  & 0.016 & 0.26 \\  \hline \hline
\end{tabular}
\end{table}

The experimental results we have presented here are in qualitative agreement with the large Peltier effect in the quantum Hall plateau regions. 
The apparent absence of the component of $\Delta C$ having the same sign for both $I_\mathrm{dc}$ directions in Fig.\ \ref{expFigs}(b) in the quantum Hall plateau regions suggests that the Joule heating does not have an appreciable effect on the temperature in these regions,  justifying the interpretation assuming the dominance of the Peltier effect in the incoming/outgoing thermal flux.
To more quantitatively relate the observed temperature change to $\Pi_{rr}$, we need to take into account all the phenomena involved in the thermal flux, including the Joule heating, diffusion to the electrodes, and electron-phonon interactions \cite{KajiokaCO13,Endo22}, in addition to the Peltier effect. Experimentally, measurements of detailed $I_\mathrm{dc}$ dependence will be of use to sort out contributions from the Peltier effect ($\propto I_\mathrm{dc}$) and the Joule heating ($\propto {I_\mathrm{dc}}^2$), while varying $T_\mathrm{bath}$ will help examining contributions from thermal diffusion, electron-phonon interaction with deformation-potential coupling, and that with piezo-electric coupling, which varies as a function of $T^2$, $T^7$, and $T^5$, respectively  \cite{KajiokaCO13,Endo22}. It is also necessary to understand in more detail the mechanism through which $C$ varies with the temperature. Such study is currently under progress and will be the subject of our future publication.
 
\section{Conclusion}\label{secConc}
We have calculated the Peltier coefficient $\Pi_{rr}$ of Corbino quantum Hall systems for various temperatures and quantum mobilities using the analytic formula deduced from the zero-temperature electrical conductivity obtained by SCBA \cite{AndoLB74} modified to incorporate spin splitting. Large negative (positive) $\Pi_{rr}$ in the $\nu \gtrsim$ integer ($\nu \lesssim$ integer) regions within the quantum Hall plateau areas is found to grow with decreasing temperature or decreasing disorder, approaching the upper limit given by Eq.\ (\ref{EqPirrCLT}). 
In an attempt to experimentally find the evidence for the large Peltier effect, we measured the temperature $T_\mathrm{out}$ near the outer perimeter of a Corbino disk while applying a radial dc current $I_\mathrm{dc}$. We found, as expected from the Peltier effect, that $T_\mathrm{out}$ becomes higher or lower than the bath temperature $T_\mathrm{bath}$ depending on the direction of the thermal flux $\Pi_{rr}I_\mathrm{dc}$ carried by the Peltier effect. The present study suggests the possibility of achieving, by employing the Peltier effect, the electron temperature lower than the temperature attainable by a dilution refrigerator.

\backmatter


\begin{appendices}
\section{Model for exchange-enhanced effective g-factor}\label{secAPXg}
\begin{figure}[h]
\centering
\includegraphics[width=0.8\textwidth]{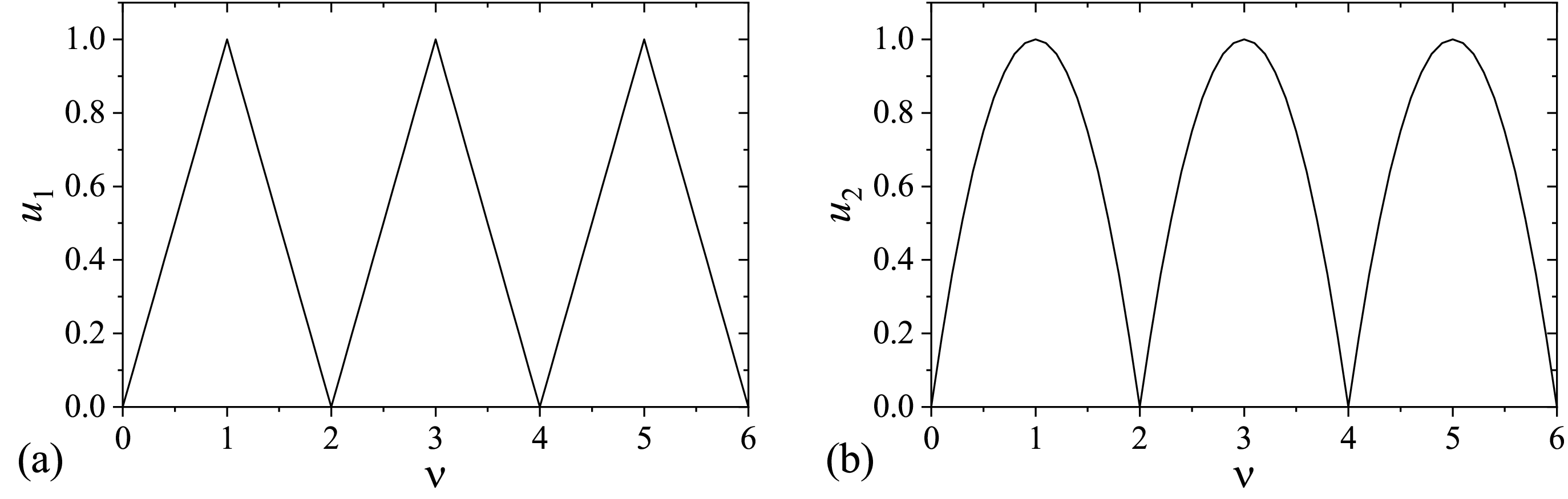}
\caption{The functions $u_1(\nu)$ (a) and $u_2(\nu)$ (b).}\label{u1u2}
\end{figure}

As mentioned in the main text, the effective g-factor $g^*$ oscillates with the Landau-level filling fraction $\nu$, taking maxima at odd-integer fillings due to the exchange-enhancement \cite{AndoOG74}. In the present study, we used a simple model function for $g^*$,
\begin{equation}
g^* = g_\mathrm{GaAs} + g_0 \exp{\left( -\frac{\pi}{\mu_\mathrm{q} |B|} \right)} u_2(\nu).  \label{geff}
\end{equation}
Here, $g_\mathrm{GaAs} = -0.44$ is the bulk g-factor for GaAs and $g_0$ represents a factor specifying the degree of the enhancement. In this study, we used $g_0 = -17.9$ to roughly reproduce the experimentally observed plateau width of odd-integer quantum Hall states at low temperatures. The exponential factor accounts for the effect of the Landau-level broadening due to disorder and
\begin{equation}
u_2 ( \nu ) = 1 - (1 - u_1( \nu ))^2 \label{u2}
\end{equation}
is the function that quadratically tend to unity when $\nu$ approaches an odd integer and becomes zero at an even-integer $\nu$ [Fig.\ \ref{u1u2}(b)], where 
\begin{equation}
u_1 (\nu ) = \frac{1}{2} + (-1)^{[ \nu ]} \left( \nu - [ \nu ] - \frac{1}{2} \right) \label{u1}
\end{equation}
represents the function that linearly connects adjacent integer $\nu$ with $u_1(\nu) = 0$ and $1$ at an even and an odd $\nu$, respectively, where $[\nu]$ represents the integer part of $\nu$ [Fig.\ \ref{u1u2}(a)].

\section{Function $F_n(x)$}\label{secAPXF}
The function $F_n(x)$ defined in Eq.\ (\ref{EqF}) can be written analytically using the polylogarithm function,
\begin{equation}
{\rm Li}_s\left(z\right)=\sum_{k=1}^{\infty}\frac{z^k}{k^s}.    \label{EqLi}
\end{equation}
Here, we present the first three of them, which are relevant to the present study:
\begin{subequations} \label{EqF1to4}
\begin{align}
F_1\left(x\right) & ={\rm Li}_1\left(-e^{-x}\right),  \label{EqF1}\\
F_2\left(x\right) & =x{\rm Li}_1\left(-e^{-x}\right)+{\rm Li}_2\left(-e^{-x}\right),  \label{EqF2}\\
F_3\left(x\right) & =x^2{\rm Li}_1\left(-e^{-x}\right)+2x{\rm Li}_2\left(-e^{-x}\right)+2{\rm Li}_3\left(-e^{-x}\right).  \label{EqF3} 
\end{align}
\end{subequations}

\section{Approximation near the center of the plateau}\label{secAPXP}
Deep in the quantum Hall plateau area near an exact integer filling, where the chemical potential is well separated in energy from two adjacent Landau levels on both sides, we have $|\xi_{N \sigma}| \gg |\eta|$ for clean 2DES having small $\Gamma$. In such regions, we can readily see from Eqs.\ (\ref{Eqsgmrr}), (\ref{Eqepsrr}), 
(\ref{EqDandA}), and Appendix \ref{secAPXF} that $\sigma_{rr}$ and $\varepsilon_{rr}$ 
become infinitesimally small at low temperatures. By contrast, $S_{rr}$ and $\Pi_{rr}$, 
the subject of our main interest in the present study,  generally own finite values despite being comprised of the ratio between such infinitesimally small values [see Eqs.\ (\ref{EqSrr}) and (\ref{KelvOns})]. 
Extremely small terms contained in $F_n(x)$ [see Eq.\ (\ref{EqF1to4})] can become problematic in actual numerical calculations of  $D_n(\xi_{N \sigma}, \eta)$. 
In order to avoid the difficulty, we introduce approximate formulas for $D_n(\xi, \eta)$ valid at $|\xi| \gg |\eta|$ in this appendix, which makes the calculations of $S_{rr}$ and $\Pi_{rr}$ 
more tractable when approaching integer fillings.

Noting that the argument $z = -e^{-(\xi \pm \eta)}$ in the polylogarithm function becomes extremely small in the region considered here, we can safely neglect higher order terms in Eq.\ (\ref{EqLi}) and simply have ${\rm Li}_s(z) \simeq z$. With this approximation, $D_n(\xi, \eta)$ for $n = 1 - 3$ can be written as
\begin{subequations}\label{ADs}
\begin{align}
D_0(\xi, \eta) & \simeq 2 e^{-|\xi|} \sinh{\eta}, \label{AD0} \\
D_1(\xi, \eta) & \simeq \mathrm{sgn} (\xi) 2e^{-|\xi|} \left[-\eta \cosh{\eta} +\left(1+|\xi|\right) \sinh{\eta} \right], \label{AD1} \\
D_2(\xi, \eta) & \simeq 2 e^{-|\xi|} \left[-2\left(1+\left|\xi\right|\right)\eta\cosh{\eta}+\left(2+2\left|\xi\right|+\xi^2+\eta^2\right)\sinh{\eta}\right], \label{AD2}\\
D_3(\xi, \eta) & \simeq \mathrm{sgn} (\xi) 2e^{-|\xi|} \left\{-\left(6+6\left|\xi\right|+3\xi^2+\eta^2\right)\eta\cosh{\eta}\right. \nonumber \\
& \hspace{5mm} +\left. \left[6+6\left|\xi\right|+3\xi^2+\left|\xi\right|^3+3\left(1+\left|\xi\right|\right)\eta^2\right]\sinh{\eta}\right\}. \label{AD3} 
\end{align}
\end{subequations}


\section{Approximation for vanishingly small disorder}\label{secAPXD}
In the limit $\eta = \Gamma / (k_\mathrm{B}T) \rightarrow 0$, we have
\begin{equation}
D_n(\xi_{N \sigma}, \eta) \simeq 
2 \eta {\xi_{N \sigma}}^n \phi \left( \xi_{N \sigma} \right),  \label{EqAppD}
\end{equation}
where we defined $\phi (x) \equiv  -\partial \tilde{f} / \partial x = [4 \cosh^2{(x/2)}]^{-1}$. With this approximation, Eqs.\ (\ref{EqsgmrrSCBA}) and (\ref{EqepsrrSCBA}) becomes
\begin{equation}
\sigma_{rr}  = \frac{e^2}{h}\frac{2}{\pi} \sum_{N \sigma}\left(N+\frac{1}{2}\right)\left(\alpha_{N \sigma} +\beta_{N \sigma}  \xi_{N \sigma}-\gamma {\xi_{N \sigma}}^2 \right) \phi \left( \xi_{N \sigma} \right)
\label{EqsgmrrC}
\end{equation}
and
\begin{equation}
\varepsilon_{rr}  = -\frac{k_B}{e}\frac{e^2}{h}\frac{2}{\pi} \sum_{N \sigma}\left(N+\frac{1}{2}\right)\left(\alpha_{N \sigma} + \beta_{N \sigma}  {\xi_{N \sigma}}-\gamma {\xi_{N \sigma}}^2 \right)  \xi_{N \sigma} \phi \left( \xi_{N \sigma} \right),
\label{EqepsrrC}
\end{equation}
respectively.
At low temperature where overlap between spin-split Landau levels can be totally neglected, we only have to consider the highest occupied level characterized by the Landau-level index
\begin{equation}
N_\mathrm{F} = \left[ \frac{\nu}{2} \right]
\end{equation}
and the spin index
\begin{equation}
\sigma_\mathrm{F} = (-1)^{[ \nu ]},
\end{equation}
where we assumed $g^* < 0$.
This allows us to further approximate $\sigma_{rr}$ and $\varepsilon_{rr}$ as
\begin{equation}
\sigma_{rr} \simeq \frac{e^2}{h}\frac{2}{\pi} \left(N_\mathrm{F}+\frac{1}{2}\right)\left(\alpha_{N_\mathrm{F}, \sigma_\mathrm{F}} +\beta_{N_\mathrm{F}, \sigma_\mathrm{F}}  \xi_{N_\mathrm{F}, \sigma_\mathrm{F}}-\gamma {\xi_{N_\mathrm{F}, \sigma_\mathrm{F}}}^2 \right) \phi \left( \xi_{N_\mathrm{F}, \sigma_\mathrm{F}} \right) 
\label{EqsgmrrCLT}
\end{equation}
and
\begin{equation}
\varepsilon_{rr} \simeq -\frac{k_B}{e}  \xi_{N_\mathrm{F}, \sigma_\mathrm{F}}  \sigma_{rr}.
\label{EqepsrrCLT}
\end{equation}
We thus have
\begin{equation}
S_{rr} \simeq -\frac{k_B}{e}  \xi_{N_\mathrm{F}, \sigma_\mathrm{F}} =   - \frac{1}{e T} (\varepsilon_{N_\mathrm{F} \sigma_\mathrm{F}}-\zeta)
\label{EqSrrCLT}
\end{equation}
and 
\begin{equation}
\Pi_{rr} \simeq  - \frac{1}{e} (\varepsilon_{N_\mathrm{F} \sigma_\mathrm{F}}-\zeta).
\label{EqPirrCLT}
\end{equation}
By further rewriting
\begin{equation}
\begin{split}
\varepsilon_{N_\mathrm{F} \sigma_\mathrm{F}}-\zeta = \left(N_\mathrm{F}+\frac{1}{2}\right)\hbar\omega_\mathrm{c} +\frac{1}{2} \sigma_\mathrm{F} g^* \mu_\mathrm{B} B-\frac{\hbar\omega_\mathrm{c}}{2}\nu \\
= \left( \left\lfloor \frac{\nu}{2} \right\rfloor-\frac{\nu}{2} + \frac{1}{2} + \frac{(-1)^{\lfloor \nu \rfloor}}{2} \frac{g^*m^*}{2m} \right) \hbar\omega_\mathrm{c},
\end{split}
\end{equation}
where $m$ represents the electron rest mass, we can find that the minimum/maximum values that $\Pi_{rr}$ can take when approaching the integer fillings $\nu$ are 
\begin{equation}
\Pi_{rr} \rightarrow \mp \frac{1}{2e} \left( 1- \frac{|g^*|m^*}{2m} \right) \hbar\omega_\mathrm{c}
\end{equation}
for even integers and
%
\begin{equation}
\Pi_{rr} \rightarrow \mp \frac{1}{2e} \left(\frac{|g^*|m^*}{2m} \right) \hbar\omega_\mathrm{c}
\end{equation}
for odd integers, where the upper (lower) sign represents approaching from higher (lower) $\nu$ side.
%

\section{Reproducibility of the change in $C$ induced by $I_\mathrm{dc}$}\label{secAPRpr}
\begin{figure}[h]
\centering
\includegraphics[width=1.0\textwidth]{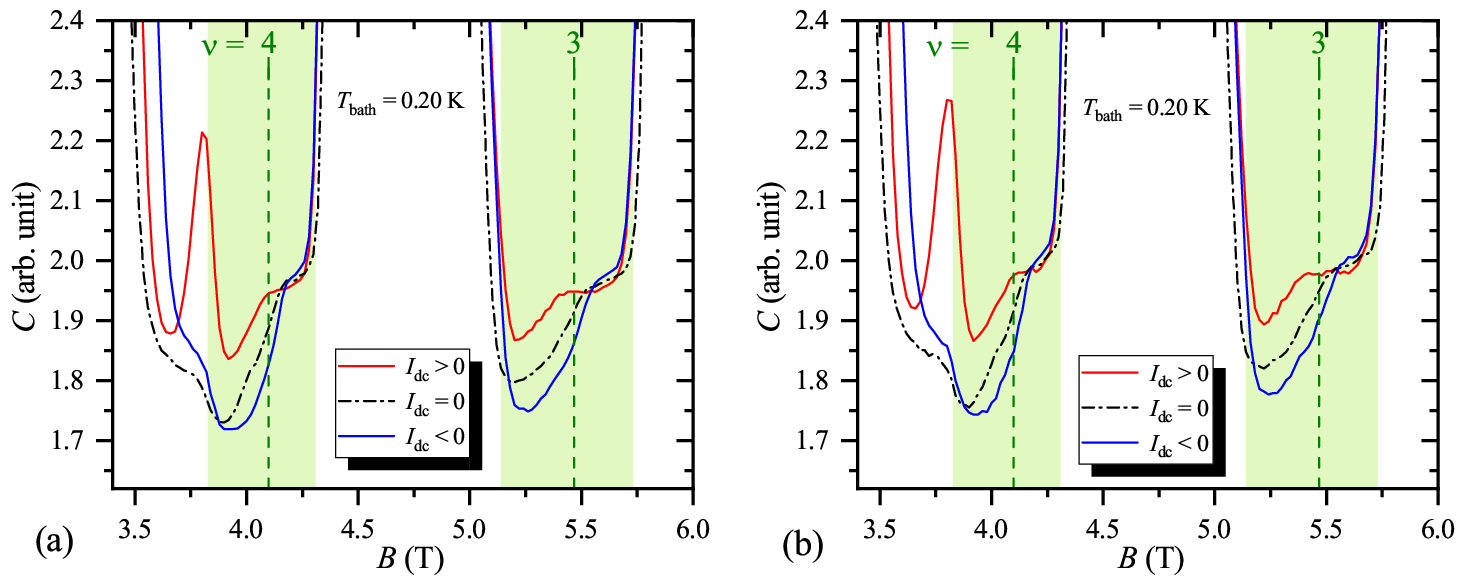}
\caption{$C$ vs.\ $B$ measured with $I_\mathrm{dc} \gtreqqless 0$. Here and in Fig.\ \ref{expFigs}(b), data for different $I_\mathrm{dc}$ were taken by first fixing the magnetic field $B$ and then varying the source-drain bias $V_\mathrm{sd}$ to change $I_\mathrm{dc}$ at the fixed $B$. Starting from $B = 6$ T, the measurements were repeated stepwise, reducing $B$ by 0.02 T in one step. (a) Measurements performed with the identical experimental setup as in Fig.\ \ref{expFigs}(b) but with reversed order of varying $I_\mathrm{dc}$. (b) Measurements performed using a different voltage source for driving $I_\mathrm{dc}$.}\label{LPandKeithley}
\end{figure}

\begin{figure}[h]
\centering
\includegraphics[width=1.0\textwidth]{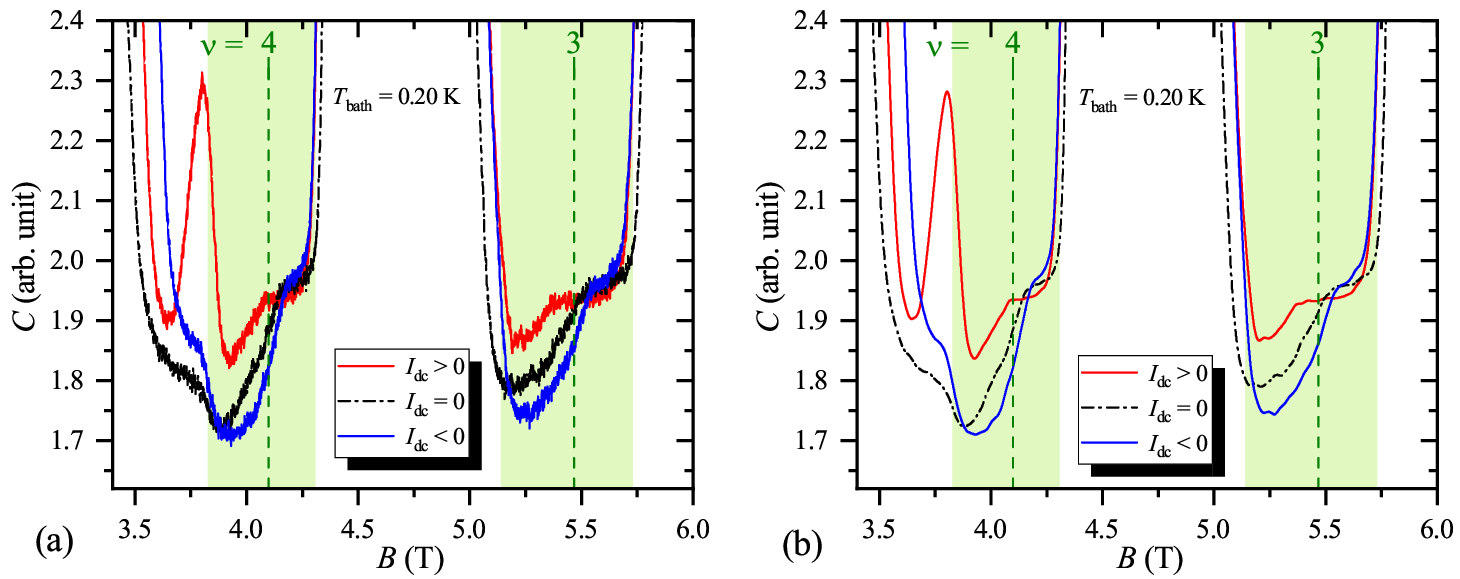}
\caption{(a) $C$ vs. $B$ obtained by sweeping $B$ (taking one data point per $\sim$5$\times$10$^{-4}$ T) at fixed $V_\mathrm{sd}$. (b) After numerical smoothing of the data shown in (a) to improve the signal-to-noise ratio. }\label{Bswps}
\end{figure}

In this appendix, we demonstrate that the $I_\mathrm{dc}$-induced change in $C$ shown in Fig.\ \ref{expFigs}(b), despite being subtle especially in the lower-$\nu$ side of the quantum Hall plateau regions, was reproducibly and robustly observed in the measurements repeated with slightly differing conditions.  

In Fig.\ \ref{expFigs}(b), the data acquisition was done by first fixing the magnetic field $B$ and changing $I_\mathrm{dc}$ at fixed $B$, in the order $I_\mathrm{dc} > 0$, $I_\mathrm{dc} = 0$,  $I_\mathrm{dc} < 0$. The procedure was repeated by changing the magnetic field by 0.02 T in one step. Suitable averaging was performed for each data point to improve the signal-to-noise ratio. We used a low-noise dc voltage source LP6016 (NF Corporation) to apply source-drain bias $V_\mathrm{sd}$ for driving $I_\mathrm{dc}$. In Fig.\ \ref{LPandKeithley}(a), we show the data taken with the identical experimental setup as in Fig.\ \ref{expFigs}(b) but reversing the order of the $I_\mathrm{dc}$ variation, starting from $I_\mathrm{dc} < 0$.  The data thus obtained were almost indistinguishable from the data shown in Fig.\ \ref{expFigs}(b). The data were found to remain roughly the same when we replaced the voltage source with Keithley 2450, as displayed in Fig.\ \ref{LPandKeithley}(b).

The data shown in Fig.\ \ref{Bswps} were taken by sweeping the magnetic field at fixed $V_\mathrm{sd}$. As can be seen in Fig.\ \ref{Bswps}(b), the data shown in Figs.\ \ref{expFigs}(b) and \ref{LPandKeithley} were roughly reproduced by taking suitable moving average to the raw data, although taken with substantially different process. The small change in the lower-$\nu$ side is already discernible even in the raw data shown in Fig.\ \ref{Bswps}(a), indicating that the change in $C$ induced by $I_\mathrm{dc}$ exceeds the noise level of the measurement of $C$.




\end{appendices}


\bmhead{Data Availability}
Data sets generated during the current study are available from the corresponding author on reasonable request.

\bmhead{Acknowledgements}
This work was supported by JSPS KAKENHI Grant Numbers 24K06914 and 23K23073.

\bibliography{ourpps,thermo,lsls,twodeg,qhe,TextBooks,spincal}

\end{document}